\documentclass{aa}  
\usepackage{todonotes}
\usepackage{graphicx}
\usepackage{natbib}
\usepackage{pdfpages}
\usepackage{color}
\usepackage[varg]{txfonts}
\usepackage{amsmath}

\def\kms {\rm{km~s^{-1}}}
\def\apj {ApJ}
\def\apjl {ApJL}
\def\apjs {ApJS}
\def\aj {AJ}
\def\aap {A\&A}
\def\mnras {MNRAS}

\def\nat {Nature}
\def\arcsec{''}

\def\pc {\rm pc}
\def\Gyr {\rm Gyr}
\def\Myr {\rm Myr}

\newcommand{\angstrom}{\textup{\AA}}

\begin{document}

   \title{Effects of environment on sSFR profiles of late-type galaxies in the CALIFA survey}


   \author{Valeria Coenda\inst{1,2}
          \and Dami\'an Mast\inst{2,3}
          \and H\'ector J. Mart\'inez\inst{1,2}
          \and Hern\'an Muriel\inst{1,2}
          \and Manuel E. Merch\'an\inst{1,2}}

   \institute{Instituto de Astronom\'ia Te\'orica y Experimental (IATE), CONICET - UNC, Laprida 854,
   X5000BGR,C\'ordoba, Argentina\\
             \and
             Observatorio Astron\'omico, Universidad Nacional de C\'ordoba, Laprida 854, X5000BGR, C\'ordoba, 
             Argentina.\\
             \and 
             Consejo de Investigaciones Cient\'{i}ficas y T\'ecnicas de la Rep\'ublica Argentina, Avda. Rivadavia 1917, C1033AAJ, CABA, Argentina.\\
              }

   \date{Received XXX, XXXX; accepted XXX, XXXX}

 
  \abstract
   {}
   {We explore the effects of environment on star formation in late-type galaxies by studying the 
   dependence of the radial profiles of specific star 
   formation rate (sSFR) on environment and the stellar mass, using a sample of 275 late-type galaxies drawn from the CALIFA survey.}
   {We consider three different discrete environments: field galaxies, galaxies in pairs, and 
   galaxies in groups, with stellar masses $9\le \log(M_{\star}/M_{\odot}) \le 12$, and compare their
   sSFR profiles across the environments.}
   {Our results suggest that the stellar mass is the main factor determining the sSFR 
   profiles of late-type      
   galaxies; the influence of AGNs and bars are secondary. We find that the relative size of    
   the bulge plays a key role in depressing star formation towards the center of late-type            
   galaxies. The group environment determines clear differences in the sSFR profiles of galaxies.      
   We find evidence of an outside-in action upon galaxies with stellar masses 
   $9\le \log(M_{\star}/M_{\odot}) \le 10$
   in groups. We find a much stronger suppression of star formation in the inner regions of massive 
   galaxies in groups, which may be an indication of a different merger history.
   }
   {}

   \keywords{Techniques: Integral Field Spectroscopy --
             Galaxies: general -- 
             Galaxies: formation --
             Galaxies: star formation -- 
             Galaxies: groups: general  
               }

   \maketitle
%
\section{Introduction}\label{intro}

One of the greatest questions surrounding galaxy formation pertains to the evolution of the baryonic component. More specifically,  the different mechanisms that can induce or quench star formation in galaxies remain poorly understood. In the local universe, galaxies have lower levels of star formation activity than in the past. The global star formation history of the universe shows a peak at $z\sim 1$ that drops to $z \sim 0$ (e.g., \citealt{Madau:1996,Sobral:2013,Khostovan:2015}). There are several processes by which galaxies can be quenched, some dependent on stellar mass and others on the environment. Different mechanisms that shut down star formation due to the intrinsic properties of the galaxy are referred to as mass quenching or internal processes. On the other hand, many quenching mechanisms are associated with the environment or "external quenching". Both, internal and environmental processes seem to affect the star formation rate (SFR) in galaxies (e.g., \citealt{Peng:2010,Sobral:2011,Muzzin:2012,Darvish:2015,Darvish:2016}).

There are a number of proposed internal processes that can remove the gas supply that fuels star formation. Studies of the evolution of the stellar mass function of star-forming and quiescent galaxies suggest that when star forming galaxies reach a stellar mass of $\sim 6\times 10^{10} M_{\odot}$ , they are quenched and become quiescent (e.g., \citealt{Peng:2010}). Among the suggested processes are halo heating \citep{Marasco:2012}, the effects of supernova(SN)-driven winds (e.g., \citealt{Stringer:2012,Bower:2012}), the feedback from massive stars (e.g., 
\citealt{DallaV:2008,Hopkins:2012}), and AGN feedback (e.g., \citealt{Nandra:2007, Hasinger:2008, Silverman:2008, Cimatti:2013}). Some authors propose AGN feedback as the primary mechanism behind both the suppression or quenching of star formation in massive galaxies, and the correlation of the central black hole
mass with the galactic bulge mass \citep{DiMatteo:2005,Martin:2005,McConnell:2013}.

Several mechanisms have been proposed to be responsible for the star-formation quenching influenced by the environment. Ram pressure stripping of the cold gas in clusters of galaxies is well established \citep{GG:1972,Abadi:1999,Book:2010,Steinhauser:2016} and can also act in less massive systems such as groups (e.g., \citealt{Rasmussen:2006,Jaffe:2012,Hess:2013})
and compact groups \citep{Rasmussen:2008}. 
However, \citet{Rasmussen:2008} found that ram pressure stripping alone could not explain the gas deficiencies in massive groups. 
Star formation can also be triggered by ram pressure in the stripped gas with
new stars tracing the gas tails \citep{Kenney:1999,Yoshida:2008,Kenney:2014}, the so called jelly-fish galaxies. 
Another mechanism that acts on galaxies in dense environments is tidal 
harassment either by the nearest neighbors, or by the  gravitational potential of the system in question \citep{Moore:1996,Moore:1999}. 
Galaxy-group interactions such as strangulation can remove warm and hot gas from a galactic halo, cutting the supply of gas for star formation \citep{Larson:1980,Kawata:2008}.
\citet{Peng:2015} argued that strangulation is the primary mechanism responsible for quenching star formation in
local galaxies, with a typical timescale of $\sim 4~ \Gyr$. Strangulation is also predicted by \citet{Kawata:2008} to act on galaxy groups.

The quenching mechanisms mentioned above act on different spatial scales and are sensitive to specific 
structural component of the galaxies. \citet{Smethurst:2015} found that quenching timescales are 
correlated with galaxy morphology. Previously, \citet{Martig:2009}  presented the concept of 
morphological quenching to analyze disk instability as the origin of the spheroidal component. Bars have also 
been proposed to reduce star formation in galaxies (e.g., \citealt{Masters:2011}).

In the last few years, thanks to the new generations of integral field spectroscopy (IFS) surveys ATLAS3D 
\citep{ATLAS3D}, CALIFA \citep{CALIFA}, SAMI \citep{SAMI}, and MANGA \citep{MANGA} it is possible to
understand star formation in galaxies in greater detail. 

Recently, \citet{Belfiore:2018} studied H$\alpha$ equivalent width and specific star formation rate (sSFR) derived from the SDSSIV-MANGA
survey \citep{MANGA,Blanton:2017}. They limited their study to galaxies with LIER emission in their
central regions (cLIERS). They found flat sSFR profiles for 
star forming and green valley galaxies with stellar masses below $\log(M_{\star}/M_{\odot})=10.5$. 
They also found that more massive star forming galaxies show an important sSFR decrease in their
central regions. They argued that this is probably a consequence of both larger bulges and an 
inside-out growth history. \citet{Spindler:2018} studied the spatial distribution of star formation of 
1494 galaxies in the local niverse from SDSSIV-MANGA and found that the sSFR of galaxies decreases 
with increasing stellar mass. In addition, they reported that massive galaxies are found to have
more centrally suppressed sSFR than low-mass galaxies, a result they related to morphology and 
the presence of AGN/LINER emission.

Using the SAMI survey, \citet{Schaefer:2017} studied the H$\alpha$ surface density gradient of 201 star forming galaxies as a function of the stellar mass and environment. Using local galaxy density as a 
characterization of the environment in which galaxies reside, they found that SFR 
gradients are steeper in dense environments for galaxies with stellar mass in the range 
$10<\log(M_{\star}/M_{\odot})<11$. This effect is accompanied by a reduction in the integrated SFR. 
The authors suggested that star formation in galaxies is suppressed with increasing local environment
density, and that this suppression starts in the outskirts of galaxies, implying that there is quenching mechanism
acting in an outside-in channel. On the other hand, \citet{Brough:2013} found no evidence of environment 
quenching on a sample of 18 galaxies studied using H$\alpha$ profiles. 

In this paper, we focus on how the sSFR of star forming galaxies is
affected by their mass and the environment in which they reside. For this purpose, we selected
late-type galaxies from the Calar Alto Legacy Integral Field Area (CALIFA) Survey \citep{CALIFA}
inhabiting three different environments: field, pairs, and groups.
The CALIFA survey is well suited to recovering the information of spatially resolved star formation in 
nearby galaxies thanks to its large 
field of view (FoV) of $74\arcsec\times64\arcsec$, which allows one to map the full optical extent of the 
galaxies up to $\sim 2.5-3$ disk effective radii, and 
the high spatial resolution of the instrument (resolving structures in galaxies at $\sim$1 kpc).
This paper is organized as follows: we describe the sample of galaxies used in Sect. \ref{data}; we provide 
details of the sSFR profiles derived from CALIFA in Sect. \ref{profiles}; we present and discuss our results
in Sect. \ref{results}; and finally, we summarize the main results of the paper in Sect. \ref{summary}.

   \begin{figure}
   \centering
   \includegraphics[width=\hsize]{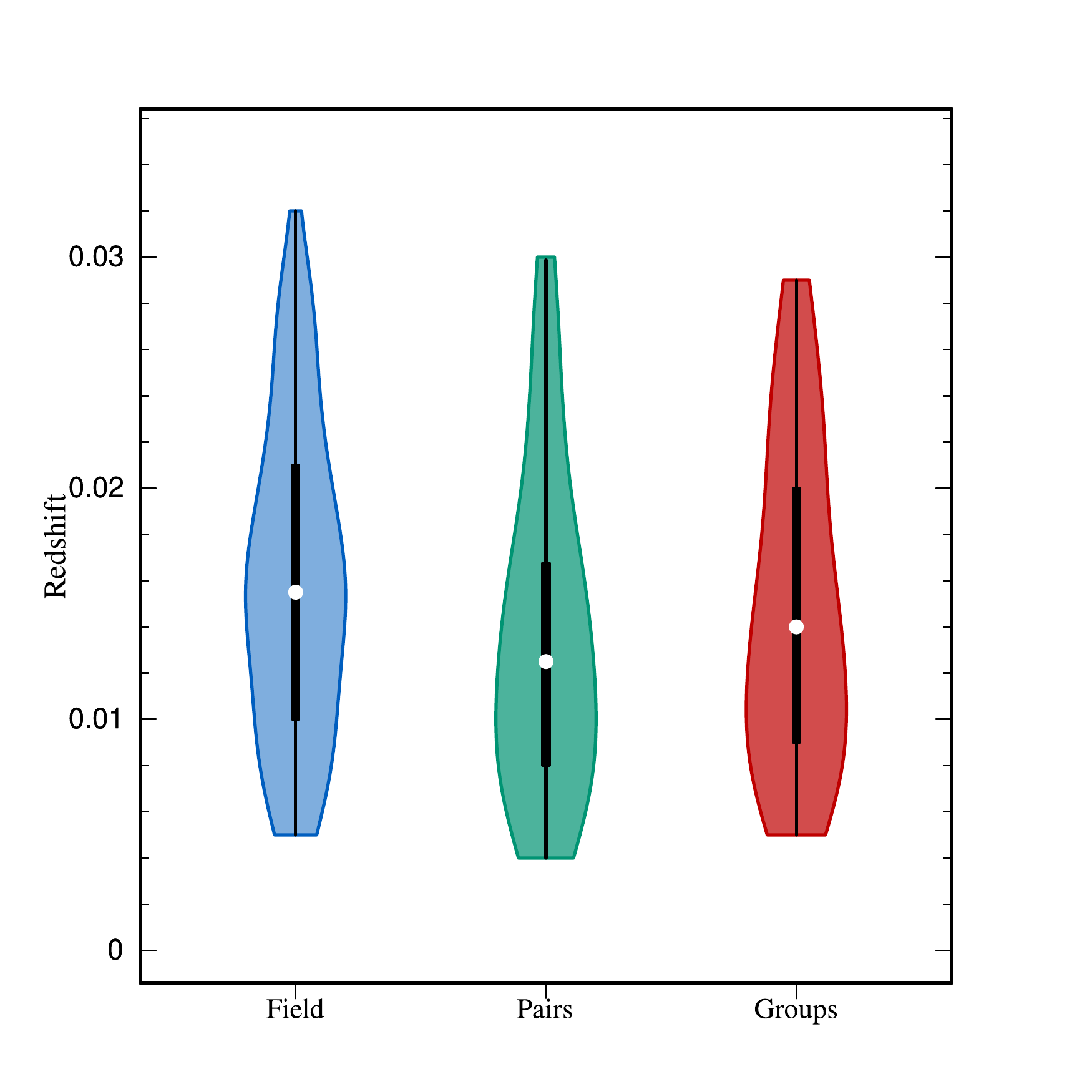}
      \caption{Violin plot of the redshift distributions of our sample of late-type galaxies. The box plot inside each violin shows the interquartile range. The inner dot in the box plot represents the median of the distribution. The widths of the violin plots are scaled by the number of observations in each bin: 96 field galaxies, 64 galaxies in pairs, and 67 galaxies in groups. Only galaxies with maps of sSFR are shown.}
         \label{fig:z}
   \end{figure}
   \begin{figure}
   \centering
   \includegraphics[width=\hsize]{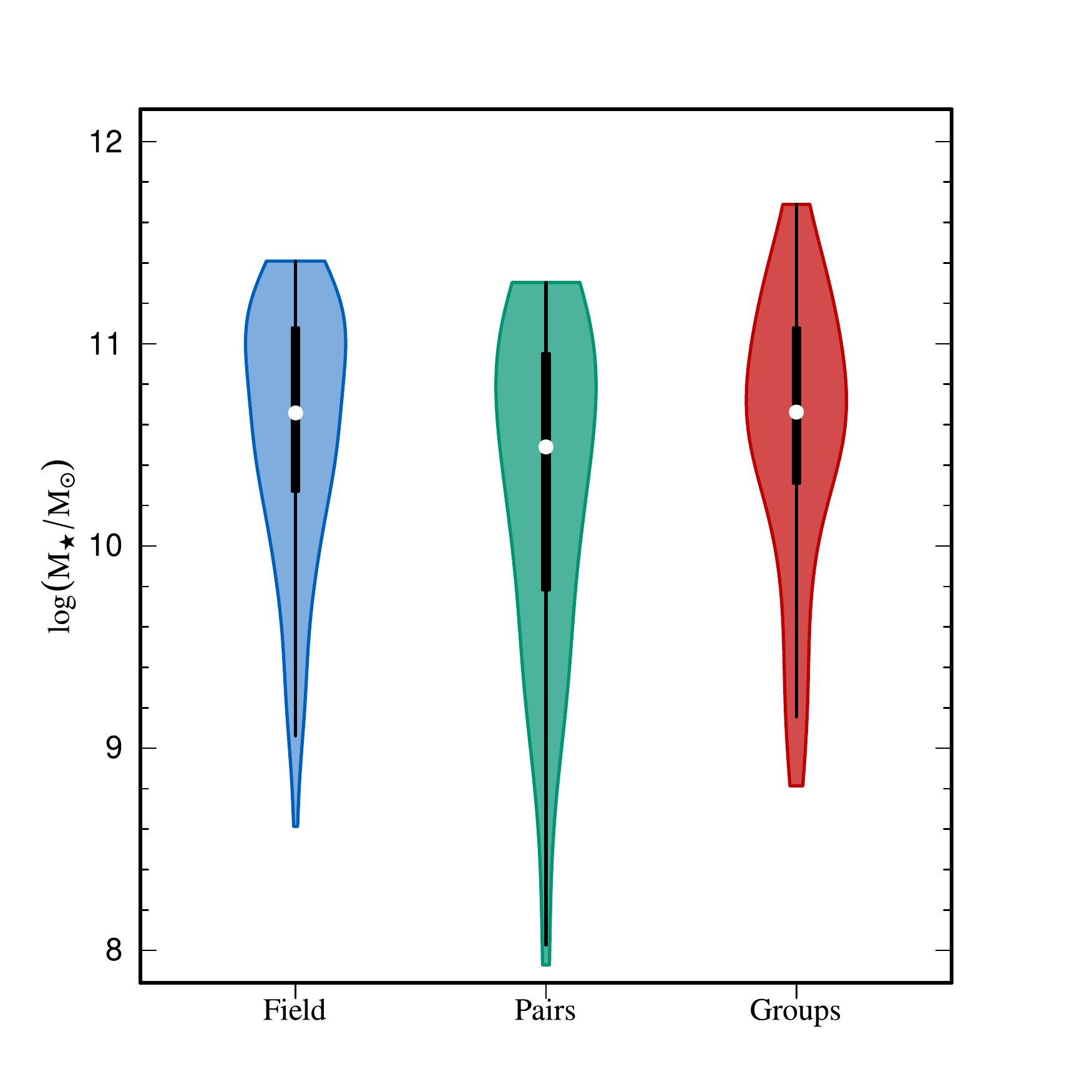}
      \caption{Violin plot of the stellar mass for late-type galaxies as a function of the environment, as in Fig. \ref{fig:z}.}
         \label{fig:mass}
   \end{figure}

 \section{The Sample}\label{data} 
 
 \subsection{The CALIFA data}
 
The Calar Alto Legacy Integral Field Area (CALIFA) survey obtained data of  science-grade quality for 667 galaxies. 
These data were made public through several data releases (DR1, \citealt{Husemann:2013}, DR2, \citealt{Garcia-Benito:2015}, 
and DR3, \citealt{Sanchez:2016}). Data were obtained with the Potsdam Multi-Aperture Spectrograph (PMAS, \citealt{Roth:2005}) 
using PPak, the integral-field spectrograph (IFU, \citealt{Kelz:2006}) mounted on the 3.5 m telescope at the Calar Alto Observatory. 
Three different spectral setups are available: I) a low-resolution V500 setup covering the wavelength range $3745-7500 \angstrom$, with a 
spectral resolution of $6.0\angstrom$, (FWHM) for 646 galaxies, II) a medium-resolution V1200 setup covering the wavelength 
range $3650-4840\angstrom$, with a spectral resolution of $2.3\angstrom$, (FWHM) for 484 galaxies, and III) a COMBO setup obtained 
from the combination of the cubes from both mentioned setups with a spectral resolution of $6.0\angstrom$, and a wavelength 
range between $3700$ and $7500\angstrom$, for 446 galaxies.

 
The CALIFA mother sample consists of galaxies selected from the SDSS DR7 (\citealt{dr7}) photometric galaxy catalog 
in the redshift range $0.005<z<0.03$, and with $r$-band angular isophotal diameter within
$45\arcsec -80\arcsec$. This sample selection spans the color-magnitude diagram and probes a wide range of stellar masses, ionization conditions, and morphological types. The galaxies were morphologically classified by five members of the collaboration 
through visual inspection of the SDSS $r$-band images\footnote{Available as ancillary data tables at
http://www.caha.es/CALIFA/ public\_html/?q=content/dr3-tables}. The sample and its characteristics are 
described in \citet{Walcher:2014}. More details on the datasets corresponding to each release can be found in the respective papers.


For the AGN identification in the sample, we consider the previously performed analyses using standard diagnostic diagrams and the WHAN diagram \citep{CidFernandes:2011}, from the central SDSS-DR7 spectra \citep{Walcher:2014} and the central spectral extractions of the CALIFA data cubes \citep{CALIFA,Husemann:2013,Singh:2013}.

In our study, we considered two different sets of data products obtained from CALIFA data cubes. Alternatively, \cite{Sanchez:2016RMxAA} used PIPE3D, an analysis pipeline based on the FIT3D fitting tool developed to explore the properties of the stellar populations and ionized gas of integral field spectroscopy (IFS) data. Also, \citealt{deAmorim:2017} used the spectral synthesis code STARLIGHT\footnote{http://starlight.ufsc.br} \citep{CidFernandes:2005} and the PyCASSO\footnote{Python CALIFA Starlight Synthesis organizer, http://pycasso.ufsc.br, mirror at http://pycasso.iaa.es} platform \citep{Cid:2013} to provide a value-added catalog of stellar population properties of 445 CALIFA galaxies (all DR3 cubes with COMBO data). Their catalog consists of maps for the stellar mass surface density, mean stellar ages and metallicities, stellar dust attenuation, SFRs, and kinematics. They also provide a catalog of integrated properties obtained from their analysis with STARLIGHT.

\cite{deAmorim:2017} presented maps obtained from fits using two sets of single stellar population (SSP) bases, labeled GMe and CBe. In the present analysis we make use of the former. In brief, base GMe is a combination of 235 SSP spectra from \cite{Vazdekis:2010} for populations older than 63 $\Myr$, and \cite{GonzalezDelgado:2005} models for younger ages. The evolutionary tracks are those from \cite{Girardi:1993} and the Geneva tracks \citep{Schaller:1992} for the youngest ages (1 and 3 $\Myr$). The Initial Mass Function is Salpeter \citep{Salpeter:1995} and the metallicity covers the seven bins of $\log(Z/Z_{\odot}) = -2.3, -1.7, -1.3, -0.7, -0.4, 0, +0.22$ \citep{Vazdekis:2010} for SSP older than 63 $\Myr$, and only the four largest metallicities for younger SSP. More details can be found in \citet{deAmorim:2017} and references therein.

In this work we concentrate on the study of the radial profile of sSFR of late-type CALIFA galaxies with stellar masses $9\le \log(M_{\star}/M_{\odot}) \le 12$, where $M_{\star}$ is the total stellar mass obtained from integrated spectra by PyCASSO \citep{Cid:2013}.
We selected all late-type galaxies 
from the final CALIFA DR3 \citep{Sanchez:2016} that have available maps of SFR and maps of stellar mass in order to construct the specific star formation maps. 
These maps are used, in turn, to 
construct sSFR radial profiles as we describe in Sect. \ref{profiles}.

\subsection{Environments}

We analyze sSFR radial profiles of late-type galaxies in three different environments: field
galaxies, galaxies in pairs, and galaxies in groups. To define these environments we use a tracer sample
of SDSS-DR12 \citep{dr12} galaxies with redshifts measured by SDSS, restricted to $r-$band Petrosian 
magnitudes $r \le 17.77$. The spectroscopic sample of the SDSS presents a redshift incompleteness 
for galaxies brighter than $r=14.5$. We improve our tracer sample by including all galaxies in 
the DR12 photometric database that have no redshift measured by SDSS, but have available redshift in 
the NED\footnote{The NASA/IPAC Extragalactic Database (NED) is operated by the Jet Propulsion Laboratory, California Institute of Technology, under contract with the National Aeronautics and Space Administration, https://ned.ipac.caltech.edu/} database. 
Given that CALIFA only observed nearby galaxies ($z<0.03$), this addition from the 
NED database ensures a high level of completeness, thus providing an adequate tracer sample to  
characterize the environment of our target sample.
  
\subsubsection{Galaxies in groups} 

Our sample of late-type galaxies in groups comprises all CALIFA galaxies that we have identified as members of groups of galaxies in our tracer sample. 
For this purpose we firstly identify groups of galaxies over this sample, using the same procedure as \citet{Merchan&Zandivarez:2005}. We refer the reader to that paper for a detailed description of the procedure. We provide here only a brief summary of the identification. Groups are identified using the algorithm developed by \citet{H&G:1982} that groups galaxies into systems using a redshift-dependent linking length. This linking length is set to recover regions with a numerical overdensity of galaxies of 200. We impose a lower limit in membership excluding groups with less than four galaxy members. Line-of-sight velocity dispersions are computed using the Gapper estimator for groups that have less than 15 members, and the bi-weight estimator for richer groups \citep{Girardi:1993,Girardi:2000}. Virial masses of these groups are computed using their velocity dispersion and projected virial radius. The sample of groups comprises 17021 groups with at least four members in the redshift range $0<z<0.3$.  Groups containing these galaxies have virial masses ranging from $\sim 1\times 10^{10} M_{\odot} $ to $ \sim 1\times 10^{16} M_{\odot} $ with a median of $\sim 8.5\times 10^{13} M_{\odot} $.
CALIFA galaxies that are found to be members of these groups amount to a total of 204, including 112 late-type galaxies.  

\subsubsection{Galaxies in pairs}

Among all CALIFA galaxies that were not identified as being part of a group, we
select those that are candidate pairs due to them having a companion inside a
line-of-sight-oriented cylinder centered in the CALIFA galaxy with a projected radius of
$100~{\rm kpc}$ and stretching $\pm 1000~\kms$ along the line
of sight \citep{Alpaslan:2015}. Our resulting sample of candidate galaxies in pairs comprises 
127 galaxies, of which 104 are late types.
To decide which of these candidate galaxy pairs are genuine pairs we proceed as follows:
firstly, we consider as center the most massive galaxy in the pair and use the relation
found by \citet{Guo:2010} to assign a halo mass to that galaxy; secondly, assuming a Navarro, Frenk \& White \citep{NFW} profile for 
the dark matter halo, we estimate its concentration parameter using 
the $z=0$ relation between concentration and halo mass by \citet{Ludlow:2014}; 
finally, we compute the escape velocity at the projected distance
between the two galaxies, and if the line-of-sight relative velocity of the pair
is smaller than this escape velocity, we consider the pair to be genuine. This criterion has
been used thoroughly in the literature (e.g., \citealt{Sales:2007}). Seventy-seven CALIFA galaxies meet this criterion, 62 of which are late type.

\subsubsection{Field galaxies}

We consider as field galaxies the remaining late-type CALIFA galaxies that were not identified as being in groups or in pairs. These
field galaxies comprise 226, of which 185 are late-type galaxies.
Some of our field galaxies may not be isolated galaxies, but actual galaxies in pairs or in groups. 
Therefore, any differences
we find below between field galaxies and galaxies in pairs or in groups could actually be
more significant. We do not expect, however, this possible contamination from
galaxies in pairs or groups to change our conclusions.

\subsubsection{Testing the environments: the fraction of late-type galaxies}

It is well known that the fraction of late-type galaxies decreases as the environment becomes denser.
In order to test our classification of environments, we compute the fraction of late-type galaxies in the field, in pairs, 
and in groups. For the latter, we split the sample into three bins of virial mass. Table \ref{table:morph} 
quotes the fraction of late-type galaxies for each environment. As a general trend, the fraction of 
late-type galaxies diminishes as we move from field to massive groups. The only exception is in the low-mass bin 
of groups of galaxies, where the fraction of late-type galaxies is comparable to that of the field. This could be an 
indication that an important fraction of low-mass groups are not actual physical systems. 
In order to improve our characterization of the group environment, we excluded from our analysis groups 
with virial mass lower than $10^{12} M_{\odot}$. Table \ref{table:morph} shows the fraction of late-type galaxies of 
the resulting sample, where it can be seen that 
the fraction of late-type galaxies in groups with virial mass $\geq 10^{12}  M_{\odot}$ is clearly smaller than for 
galaxies in pairs or in the field. With this cut in the virial mass, our sample of total (late-type) CALIFA 
galaxies in groups is 180 (92). It should be noted that the size of our sample of late-type galaxies in 
groups is not large enough to be split  into bins of virial mass.

\begin{table}
\caption{Fraction of late-type galaxies as a function of the environment}
\label{table:morph}
\centering                         
\begin{tabular}{lc}        
\hline\hline                 
Environments & Fraction of \\  
             & late-type galaxies \\  
\hline
Field & $0.82$ \\
Pairs & $0.78$ \\
Groups: & \\
Low mass ($M_{vir} < 10^{12} M_{\odot}$) & $0.83$ \\
Intermediate mass ($ 10^{12} M_{\odot} \leq M_{vir} < 10^{13} M_{\odot}$) & $0.63$ \\
High mass ($M_{vir} \geq 10^{13} M_{\odot}$) & $0.31$ \\
$M_{vir} \geq 10^{12}  M_{\odot}$ & $0.52$ \\ 
\hline
\end{tabular}
\end{table}

   \begin{figure*}
   \centering
   \includegraphics[width=\hsize]{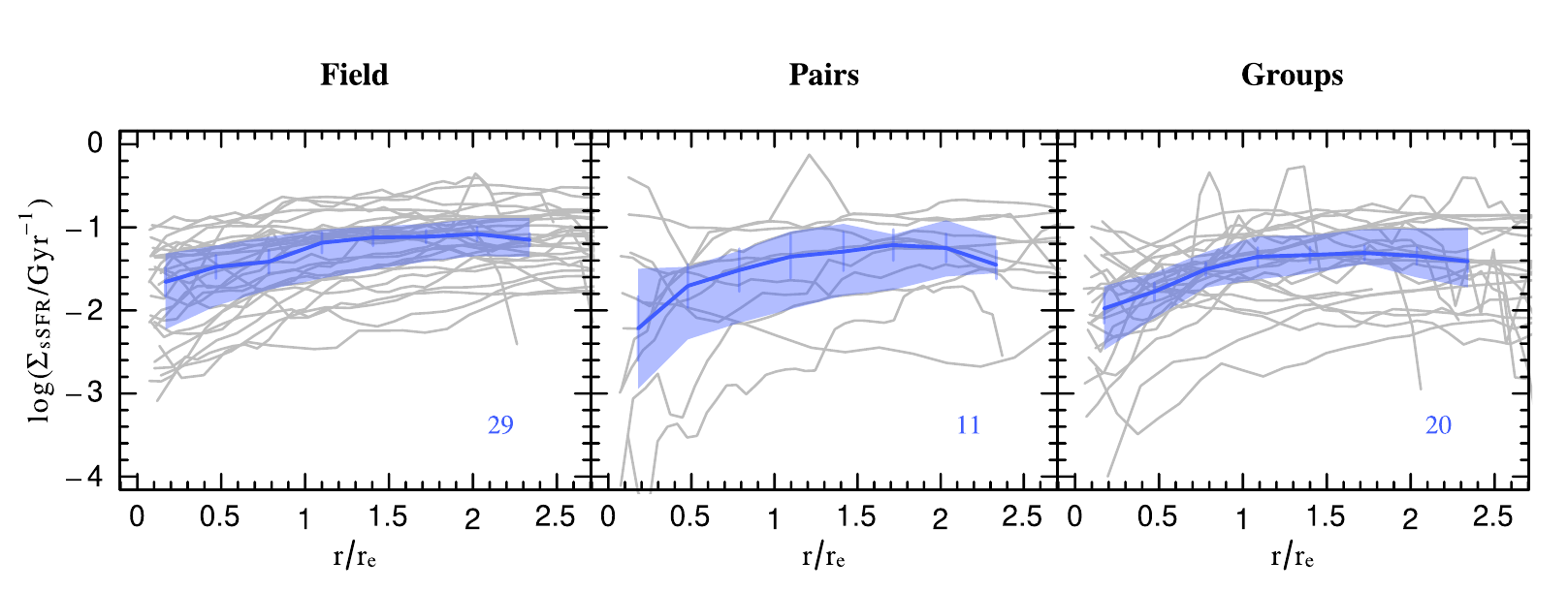}
      \caption{An example of our stacking procedure: the stacked sSFR profiles, $\Sigma_{\rm sSFR}$, for late-type galaxies in the stellar mass range $\log(M_{\star}/M_{\odot})=10.5-11.0$ in field galaxies (left panel), galaxies in pairs (central panel), and galaxies in groups (right panel). Radial distances are normalized to the half-light effective radius $r_{\rm e}$. Lines represent the median in each radial bin, and the shaded region shows the $25\%$ and $75\%$ percentiles. Vertical bars are the error in the median, and were computed by the bootstrap resampling technique. Individual profiles are shown by gray lines.}
         \label{fig:profile}
   \end{figure*}

\subsubsection{Comparison with previous analyses}

\citet{Walcher:2014} analyzed the environment associated to CALIFA's mother sample. They cross-correlated 
the mother
sample with
well-known structures, including clusters, groups, triplets, pairs, and isolated galaxies. Based on a large 
sample of known catalogs of groups and clusters, they found that between 24\% and 74\% of the galaxies in 
the mother
sample belong to a known association. This variation is due to the different membership 
criteria that were applied. For our sample, the fraction of CALIFA galaxies in groups is 34\%. For those 
galaxies belonging to a known association, \citet{Walcher:2014} separated the sample into galaxy aggregates 
with velocity dispersions $\sigma \leq 550~ \kms$, and $\sigma > 550~ \kms$. They found that 82\% of the galaxies in associations 
belong to low-velocity systems. In our sample, the percentage of galaxies in low-mass systems is 96\%. This 
difference could be related to differences in the procedure to assign members, which is more restrictive in 
our sample, thus producing lower values of the velocity dispersion. 
In addition, \citet{Walcher:2014} computed the local density around each galaxy and concluded that 
CALIFA samples all environments.

Due to the fact that we have not applied an isolation criterion, our sample of field galaxies 
can not be 
compared with the AMIGA sample of isolated galaxies \citep{Verdes:2005} used in \citet{Walcher:2014}. These 
authors found that 7\% of the mother
sample galaxies belong to isolated pairs. In our sample, pairs represent 13\%. This 
excess of pairs can be explained by the fact that CALIFA DR3 includes some extra galaxies in pairs which 
were not included in the mother
sample since they do not meet the original selection criteria.

 \section{Radial profiles}\label{profiles}
 
For the determination of radial profiles, firstly we fit ellipses to the luminosity surface density maps ($\mathcal{L}_{5635\angstrom}$). 
The construction of these maps is done by directly measuring the average flux of the spectra in the spectral window of $(5635\pm45)\angstrom$,
\citep{deAmorim:2017}. 

Using the task \textit{ellipse} \citep{Jed:1987} within IRAF\footnote{http://iraf.noao.edu/} with 1 spaxel step ($1\arcsec$) we obtain 
the ellipses that will allow us to obtain the sSFR radial profiles. The ellipse fitting was done by keeping the center of the ellipses fixed and leaving as free parameters the position angle (PA) and the ellipticity ($\epsilon$). For the center ($x_0$, $y_0$), the nuclear centroid determined from the synthetic images extracted from the CALIFA cubes was used. When the determined luminosity profiles showed anomalous behaviors (or the fit diverged), a manual ellipse fitting was carried out, controlling the variation of both PA and $\epsilon$ along the galaxy.

We then used the stellar mass surface density maps ($\Sigma_{\star}$) provided by \cite{deAmorim:2017}, which were calculated using the masses derived from STARLIGHT. These maps are in units of $M_{\odot} \pc^{-2}$. The mass has been corrected for mass that returned to the interstellar medium during stellar evolution, so it represents the mass currently trapped in stars.

Finally, we made use of the maps of mean recent SFR surface density ($\Sigma_{\rm SFR}$), which were obtained by adding the mass that became stars in the last 32 $\Myr$ and dividing this amount by this timescale \citep{Asari:2007,CidFernandes:2015}. These maps are in units of $M_{\odot} \Gyr^{-1} \pc^{-2}$.

The construction of the sSFR maps is done by dividing the stellar mass surface density maps and the SFR
surface density maps ($\Sigma_{\rm sSFR}=\Sigma_{\rm SFR}/\Sigma{\star}$). Running again the $ ellipse$ task on these maps, 
now using the ellipses obtained from the luminosity surface density maps ($\mathcal{L}_{5635\angstrom}$) fitting, we built the sSFR 
radial profiles set.

Our final sample of late-type galaxies with radial sSFR maps comprises
54 galaxies in groups, 36 galaxies in pairs, and 102 field galaxies.
We show in Fig. \ref{fig:z} the redshift distributions of our samples of spiral galaxies for each environment. As seen from the median value of each bin, which is represented by a dot in the box plot inside the violin plot, there are no significant differences in redshift distributions, which we confirm by means of Kolmogorov-Smirnov tests.

\section{Results}\label{results}
 
   \begin{figure*}
   \centering
   \includegraphics[width=\hsize]{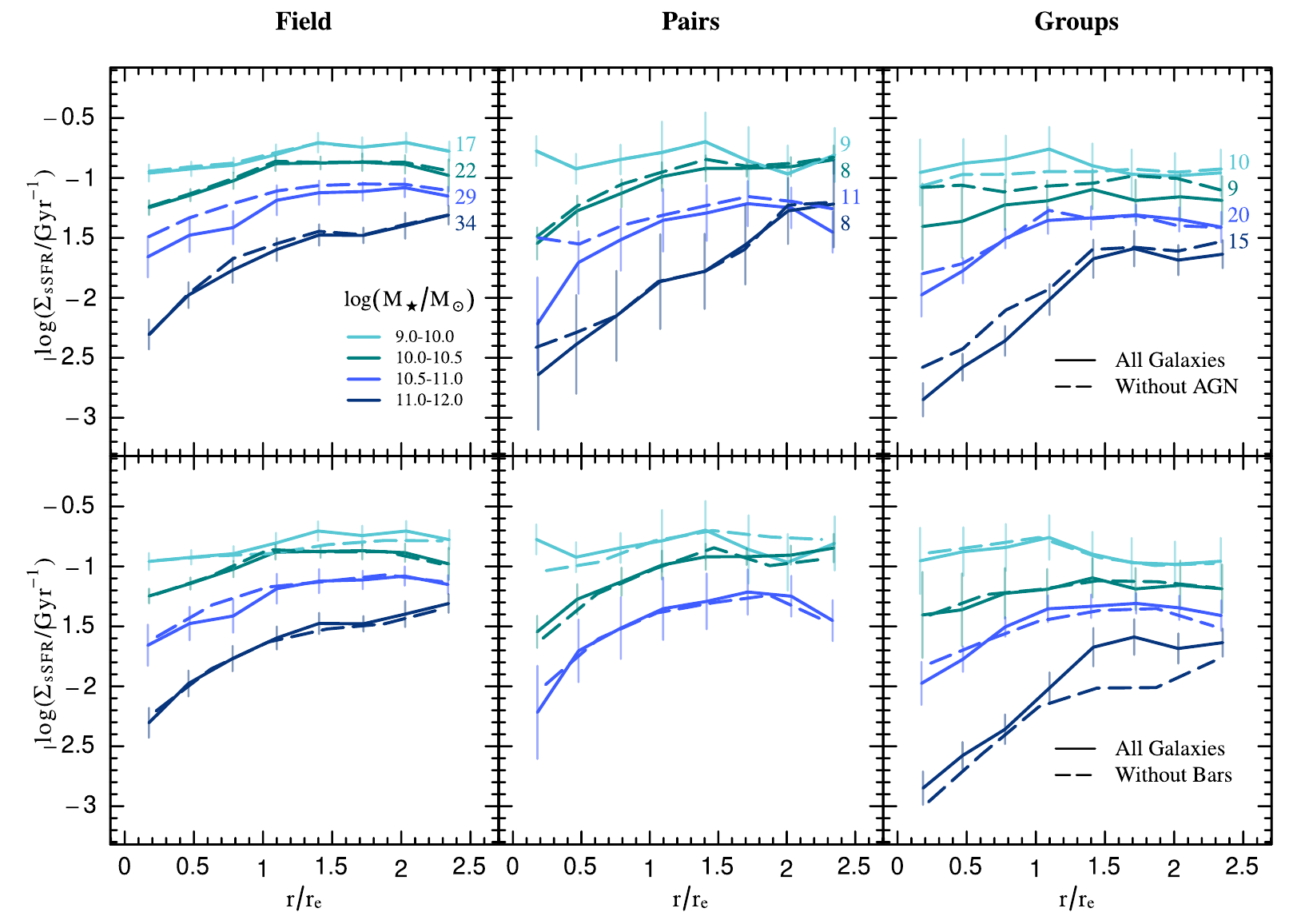}
      \caption{The median profile of the sSFR, $\Sigma_{\rm sSFR}$, for
      late-type galaxies for different mass bins as a function of the environment: field galaxies (left panel),
      galaxies in pairs (central panel), and galaxies in groups (right panel). Solid lines represent the median in
      each radial bin. Vertical error bars were computed using the bootstrap re-sampling technique. The number 
      of galaxies contributing to each median profile is quoted. Dashed lines in the upper (lower) panels
      show the median $\Sigma_{\rm sSFR}$ profiles of spirals galaxies without AGNs (bars). 
      For clarity, we excluded the vertical 
      error bars in these profiles.}
         \label{fig:radial}
   \end{figure*}
 
In this work we study the effects of external and internal mechanisms on the radial distribution of the sSFR 
for late-type galaxies. To analyze the internal processes, we investigate the sSFR profiles of late-type galaxies split into bins 
of stellar mass. To investigate the external quenching effects, we compare late-type galaxies in three discrete environments:
field galaxies, galaxies in pairs, and galaxies in groups. Our sample includes all CALIFA late-type galaxies with COMBO setup, for which we were able to construct sSFR maps. 

Figure \ref{fig:mass} shows the stellar mass distribution of late-type galaxies as a function of the environment. Median values (inner white dots inside the violin plots) are
$\log(M_{\star}/M_{\odot})=10.65$, $10.49$ and $10.66$ for groups, pairs, and field galaxies, respectively. 

We split galaxies into four stellar mass bins: $\log(M_{\star}/M_{\odot})=9.0-10.0$, $10.0-10.5$, 
$10.5.-11.0$ and $11.0-12.0$. We have re-scaled each radial profile in terms of the $r$-band 
half-light effective radius, 
$r_e$, which has been computed following \citet{Graham:2005} by using the SDSS $r-$band radius that encloses 
half the Petrosian flux and the concentration parameter in the same band. We consider the range 
$0-2.5 r_{\rm e}$ to stack the radial profiles into eight bins, 
and we calculate the median value of $r/r_{\rm e}$ for each interval of size. 
Figure 3 shows examples of sSFR radial profiles, $\Sigma_{\rm sSFR}$, for galaxies in our sample. We show the median radial profiles for galaxies in the mass bin 
$\log(M_{*}/M_{\odot})=10.5-11.0$: field galaxies (left panel), galaxies in pairs (central panel), 
and galaxies in groups (right panel). Profiles of individual galaxies are shown in gray, with the median 
profile for each bin in light blue. The shaded regions represent the $25\%$ 
and $75\%$ percentiles. Vertical error-bars were computed using the bootstrap re-sampling technique. 
Analogously to \citet{GonzalezDelgado:2016}, we have considered only mass bins containing more than five 
galaxies. We quote in all cases the number of galaxies contributing to each profile.

Figure \ref{fig:radial} compares the medians of $\Sigma_{\rm sSFR}$ as a function of mass for
each environment. In general, we observe two trends of $\Sigma_{\rm sSFR}$ with increasing stellar mass: on the one
hand, there is a global sSFR decrease, and on the other, there is a stronger drop towards the central regions.
Profiles are nearly flat for galaxies with stellar masses below $\sim 10^{10}M_{\odot}$. For more massive galaxies, 
the central suppression of the sSFR strengthens with increasing mass. This behavior is observed independently of
the environment. Our results are in agreement with \citet{Catalan-Torrecilla:2017}, \citet{Belfiore:2018}, and 
\citet{Spindler:2018} in the sense that they found flat sSFR profiles for low-mass galaxies, and a  significant 
decrease in the central regions of massive galaxies. This is consistent with the inside-out 
growth scenario or larger bulges.

There are several authors that support the inside-out mode in late-type galaxies (e.g., 
\citealt{Munoz:2007,Munoz:2011,Perez:2013,Pezzulli:2015,Ibarra-Medel:2016}). Alternatively, \citet{Abramson:2014} argue in favor of an increase in bulge mass as a result of the 
suppression in the SFR, and \citet{Belfiore:2018} argue that the sSFR suppression is not simply due to the  larger mass of the bulge, but also the innate evolution of the sSFR in the context of inside-out growth. In addition, \citet{Spindler:2018} find that at all masses, AGN/LINER galaxies are more likely to have centrally suppressed sSFR profiles than galaxies without AGN feedback.

\begin{table}
\caption{Probability of rejection of the null hypothesis in which the different
subsamples are drawn from the same underlying distribution: comparison at
fixed environment and mass bin. Each subsample is compared with its own subset resulting from the exclusion of AGNs (upper
panels of Fig. \ref{fig:radial}), or the exclusion of barred galaxies (lower panels of Fig. \ref{fig:radial}).}    
\label{table:1}
\centering                         
\begin{tabular}{ccc}        
\hline\hline                 
Mass bin & \multicolumn{2}{c}{Probability of rejection} \\
         &   No AGNs   & No Bars \\
\hline
\multicolumn{3}{c}{Field}\\
\hline
   1 & 0.67 & 0.58 \\      
   2 & 0.64 & 0.74 \\
   3 & 0.87 & 0.61 \\
   4 & 0.72 & 0.51 \\
\hline
\multicolumn{3}{c}{Pairs}\\
\hline
   1 & 0.52 & 0.63 \\      
   2 & 0.71 & 0.61 \\
   3 & 0.83 & 0.54 \\
   4 & 0.50 & $-$ \\
\hline
\multicolumn{3}{c}{Groups}\\
\hline
   1 & 0.50 & 0.64 \\      
   2 & 0.78 & 0.50 \\
   3 & 0.68 & 0.75 \\
   4 & 0.70 & 0.91 \\
\hline
\hline
\end{tabular}
\end{table}
   \begin{figure}
   \centering
\includegraphics[width=\hsize]{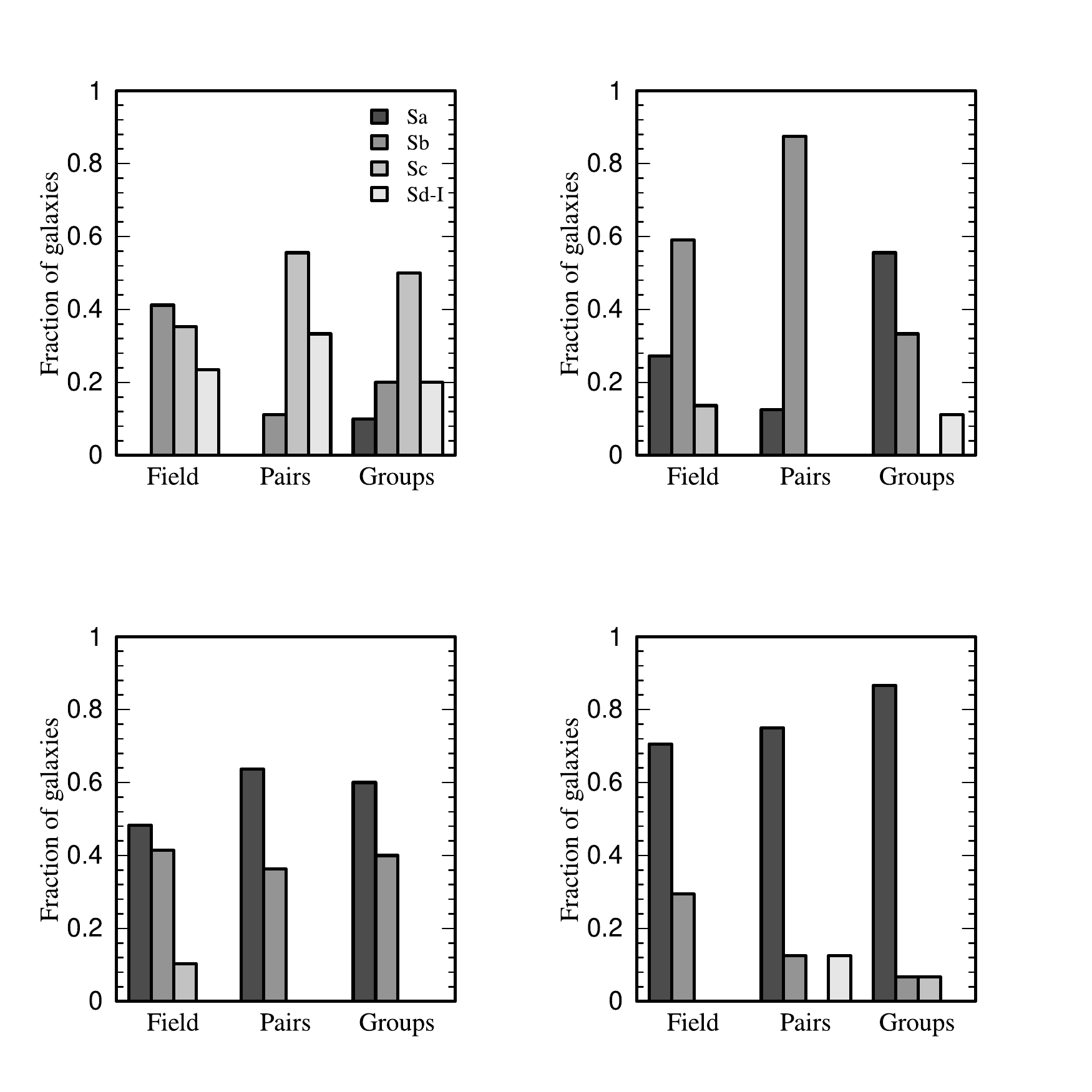}
      \caption{Comparison of the distribution of the Hubble types for our sample of late-type galaxies, as a function of the environment. From top-left to bottom-right: 
      $\log(M_{\star}/M_{\odot})=9.0-10.0$, $10.0-10.5$, $10.5.-11.0$ and $11.0-12.0$.
      } \label{fig:mor}
   \end{figure}
   \begin{figure*}
   \centering
   \includegraphics[width=15 cm]{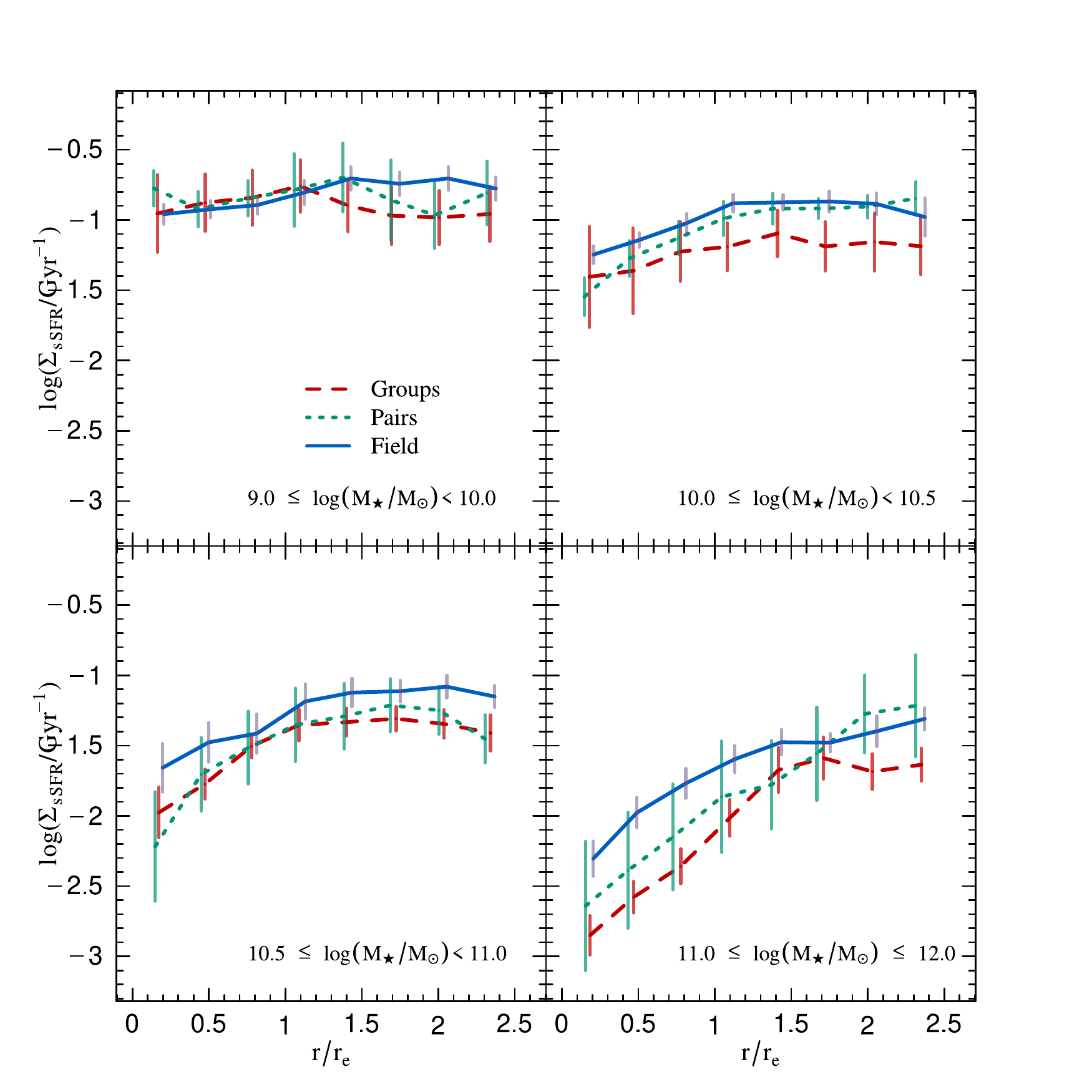}
      \caption{The stacked profile of the sSFR $\Sigma_{\rm sSFR}$ for late-type galaxies as a function of the environment. Each panel corresponds to a different mass bin. Galaxies in groups are shown in red lines, galaxies in pairs in green lines, and field galaxies in blue lines. Lines represent the median in each radial bin. Vertical error bars are as in Fig. \ref{fig:radial}. Field galaxies and galaxies in pairs have been shifted by 0.03 and -0.03 on the x-axis, respectively.}
         \label{fig:radial_env}
   \end{figure*}

We explore whether AGN feedback is the main mechanism responsible for the sSFR suppression in the inner regions of late-type galaxies. We recompute the $\Sigma_{\rm sSFR}$ radial profile of late-type galaxies excluding AGNs. This
is also seen in the upper panels of Fig. \ref{fig:radial}, as dashed lines. For the sake of clarity, we exclude the 
vertical error bars for the sample without AGNs. In general, we do not find significant differences in the 
median values of the sSFR profiles with or without AGNs. The frequencies of AGNs in the three environments 
are: 16\%, 22\%, and 35\%, for field, pair, and group galaxies, respectively. To check whether the minor 
differences seen in Fig. \ref{fig:radial} are indicative of actual differences between the populations, we rely 
on the test used by \citet{Muriel:2014} and \citet{Martinez:2016}. Applied to our problem, this test computes 
the cumulative differences between two samples in the sSFR along the radial domain, and then checks whether the 
resulting quantity is consistent with the null hypothesis in which the two samples are drawn from the same 
underlying population in regards to their sSFR profiles. 
Results of the test are quoted in Table \ref{table:1}. 
In general, the null hypothesis in which the samples including or excluding AGNs are 
indistinguishable cannot be ruled out with a high level of confidence, highlighting the role played 
by AGNs secondary to the central role of stellar mass. Probable exceptions may be the third mass bin of 
field galaxies and of galaxies in pairs, for which the largest rejection levels are reached: 87\% and
83\%, respectively.

Another factor that can play an important role in shaping the sSFR profiles is the presence of bars. 
Bars can trigger star formation in the central regions of galaxies.
Several studies have found evidence of enhanced star formation activity in barred galaxies compared to
those without bars (e.g., \citealt{Heckman:1980,Regan:2006,Lin:2017}).
On the other hand, several other works in the literature do not find such differences
(e.g., \citealt{Pompea:1990,Chapelon:1999,Willett:2015}).
Furthermore, \citet{Kim:2017} found evidence that the star formation activity of 
strongly barred galaxies is on average lower than that of galaxies without bars.
Barred galaxies amount to 34\%, 36\%, and 43\%, in our field, pairs, and group samples, respectively.
Similar to  what we have done with AGNs, we recompute the $\Sigma_{\rm sSFR}$ radial profile of late-type 
galaxies but now excluding barred galaxies. We show the resulting median profiles in the bottom panels of Fig. 
\ref{fig:radial}. We exclude the highest mass bin of galaxies in pairs, since there remained less than five 
galaxies in this bin once we had removed the barred ones. In general, the effect of bars in the radial profile of 
sSFR is negligible in our samples. The exception is the highest mass bin of galaxies in groups, where
bars appear to enhance the sSFR of these galaxies at all scales, but more notably outside 
their half-light radii.
The corresponding values of the rejection probability of the null hypothesis in which the samples 
including/excluding barred galaxies are drawn from the same population are also quoted in Table \ref{table:1}.
These rejection probability values discard bars as an important factor shaping the sSFR profiles
in late-type galaxies, with the exception of the highest mass bin in groups, for which
the null hypothesis is rejected at a 91\% level.

At this point we explore whether morphology can be an explanation for the suppression of 
star formation seen in Fig. \ref{fig:radial}, in the form of a dominant spheroidal component in the highest 
stellar mass bin. In Fig. \ref{fig:mor} we show the distributions of the Hubble types of our sample of 
late-type galaxies, as a function of stellar mass and environment. For the two lowest stellar mass bins (top 
panels), we observe a mix of Hubble types. As stellar mass increases, we observe a higher fraction of Sa and Sb 
galaxies (bottom panels). However, there is no significant environmental dependence of the morphological type 
mix. The mass dependence of the sSFR profiles 
reflects the mass dependence mix of morphologies.

In Fig. \ref{fig:radial_env} we show the same radial profiles shown in Fig. 
\ref{fig:radial} but comparing galaxies of the same mass across the three environments. 
As an overall trend, at fixed stellar mass, 
galaxies in groups are more suppressed at all radii than galaxies in pairs and in the field, with the exception
of the inner regions ($r/r_{\rm e} \lesssim 1 $) of galaxies in the lowest mass bin where there is no 
difference between environments. In general, galaxies in pairs show radial profiles that are intermediate between 
field and group galaxies, although their error bars are the largest given their small numbers.
At the lowest mass bin, galaxies in groups show interesting features. Their sSFR profile is 
consistent with that of field galaxies up to $r/r_{\rm e} \sim 1$. This contrasts with the remaining 
mass bins where group galaxies show consistently lower sSFR values at all radii. On the other hand, at 
larger radii, $r/r_{\rm e}> 1$, their star formation efficiency decays while field galaxies exhibit a flat
behavior. At the other extreme in mass, the largest differences between the three environments are seen, with
group galaxies having the steepest profile in the inner regions. 
For the two intermediate-mass bins, group and field galaxies have near parallel $\Sigma_{\rm sSFR}$ 
profiles. The (very) small sample sizes prevent us from drawing strong conclusions from galaxies
in pairs besides their sSFR intermediate between those of field and group galaxies. It is also worth 
mentioning that 87\% of our galaxies in pairs are the brightest member of their pair.
Our results do not change if we exclude those galaxies that are not the brightest member of their pair. 
In Table \ref{table:2} we show the results of the test used above, now comparing the three environments at 
fixed mass bin. For all mass bins, the highest levels of rejection (95\% level or higher) are 
reached in the comparison between field and group galaxies.

The main difference between groups and the other environments explored here is an overall decrement of the sSFR. This decrement appears to be independent of the spatial scale at least for $10 \leq \log(M_{\star}/M_{\odot}) \leq 11$. This fact could suggest that strangulation (e.g., \citealt{Larson:1980,Kawata:2008,Peng:2015}) is the main mechanism responsible for the observed overall decrement in star formation, since it is proposed to produce a uniform suppression across the galaxy \citep{vandenbergh:1991,Elmegreen:2002}. However, our sample may not cover the outermost parts of galaxies, where distinctive traces of such mechanisms may be clearer.

The differences seen in the lowest mass bin of Fig. \ref{fig:radial_env}, where we find an 
indication that the sSFR profile of group galaxies differs from that of field galaxies in a particular way,
may be an indication of the effective action over these galaxies of mechanisms that act in an outside-in mode. 
Ram pressure stripping \citep{GG:1972,Abadi:1999,Book:2010,Steinhauser:2016} could produce such a decrease of 
star formation in the outer parts of the disk, and a concentration of star formation in the inner parts 
\citep{Rasmussen:2006,Koopmann:2004a,Koopmann:2004b,Cortese:2011}, thus
providing an explanation for the observed profile. An alternative scenario could exist where tidal 
interactions produce central enhancements of star formation (e.g., \citealt{Moreno:2015}) in galaxies where global star formation is otherwise suppressed by other precesses such as
strangulation. Using numerical simulations, \citet{Yozin:2015} found that the star formation in 
galaxies in groups may be influenced by both tidal interactions leading to central enhancement of star formation and ram pressure stripping inhibiting the SFR at large radii.

On the other hand, the main differences regarding the suppression in the inner regions of 
late-type galaxies are observed only for galaxies more massive than $\log(M_{\star}/M_{\odot})=11$, where the drop becomes stronger moving from field to group galaxies. Consistently with this, \citet{Welikala:2008} found that local environment tends to reduce the SFR in central regions of galaxies.
A distinct merger history of galaxies could explain the stronger drop seen in the inner regions of massive late types in groups (bottom-right panel of Fig. \ref{fig:radial_env}); a higher frequency of mergers in the past (e.g., \citealt{Zandivarez:2006,Zandivarez:2011}) may have caused the bulges of massive group spirals to shut down their star 
formation more effectively than in pairs and in the field.

Our results are in agreement with \citet{Schaefer:2017}, although they argue that environmental quenching occurs in an outside-in mode. It is worth noting that the sample of these latter authors includes galaxies of lower mass than ours, and it is at our lowest mass bin that we find evidence of outside-in environmental effects. Another difference
with this work is that our sample extends further at the high-mass end.
On the other hand, \citet{Spindler:2018} did not find a correlation between local
environment density and the profiles of the SFR surface density. One key difference between both of these works and ours, is that here we consider discrete environments instead of a continuously parametrized measure of the environment.

\begin{table}
\caption{Probability of rejection of the null hypothesis in which the different subsamples
are drawn from the same underlying distribution: comparison between different environments at fixed stellar mass (see Fig. \ref{fig:radial_env})}    
\label{table:2}
\centering                         
\begin{tabular}{lc}        
\hline\hline                 
Environments & Probability of rejection \\  
\hline
\multicolumn{2}{c}{Mass bin 1}\\
\hline
   Field$-$Pairs  & 0.51 \\      
   Field$-$Groups & 0.96 \\
   Pairs$-$Groups & 0.86 \\
   \hline
\multicolumn{2}{c}{Mass bin 2}\\
\hline
   Field$-$Pairs  & 0.70 \\      
   Field$-$Groups & 0.95 \\
   Pairs$-$Groups & 0.83 \\
\hline
\multicolumn{2}{c}{Mass bin 3}\\
\hline
   Field$-$Pairs  & 0.95 \\      
   Field$-$Groups & 0.95 \\
   Pairs$-$Groups & 0.70 \\
\hline
\multicolumn{2}{c}{Mass bin 4}\\
\hline
   Field$-$Pairs  & 0.85 \\      
   Field$-$Groups & $>0.99$ \\
   Pairs$-$Groups & 0.80 \\
\hline
\hline
\end{tabular}
\end{table}
  
\section{Summary}\label{summary}

In this paper we analyze the dependence of the radial profiles of sSFR on environment and stellar mass, using a sample of late-type galaxies drawn from the CALIFA survey. We consider three different discrete environments: field galaxies, galaxies in pairs, and galaxies in groups. We split our galaxy samples into four bins of stellar mass, covering the range $9 \le \log(M_{\star}/M_{\odot}) \le 12$, and compare their sSFR as a function of radius (in units of their effective radius) throughout the three environments. We note the following findings.
\begin{itemize}
\item As in previous works in the literature, when we move from less to more massive
galaxies, profiles of sSFR lower and bend towards the inner regions of the galaxies.
Mass primarily determines the sSFR profiles of late-type galaxies.
\item Morphology matters: the relative size of the bulge plays a key role in depressing
star formation towards the center of late-type galaxies.
\item The influence of AGNs and bars is secondary to mass.
\item The group environment determines clear differences in the sSFR profiles of galaxies.
On the other hand, galaxies in pairs show sSFR that is intermediate between that of field and group galaxies. 
\item There is evidence of an outside-in action upon galaxies with stellar masses 
$9\le\log(M_{\star}/M_{\odot})\le10$ in groups. 
\item There is evidence of a much stronger star formation suppression in the inner regions
of massive galaxies in groups, which may be an indication of a different merger history.
\end{itemize}

To distinguish whether the flat sSFR profiles seen for low-mass late-type galaxies in groups are 
the result of ram-pressure stripping that decreases star formation in the outer parts of disks, or are due to tidal interactions that enhance central star formation, a careful analysis of the 
age of the stellar populations as a function of radius may provide further clues. In a forthcoming paper
(Coenda et al. in prep.) we focus on age profiles and explore the differences among the different environments.

\begin{acknowledgements}
This paper is based on data obtained by the CALIFA survey
(http://califa.caha.es) which is based on observations collected at the Centro Astron\'omico Hispano Alem\'an (CAHA) at Calar Alto, operated jointly by the Max-Planck-Institut f\"ur Astronomie and the Instituto de Astrof\'{i}sica de Andaluc\'{i}a (CSIC). 
This research has made use of the NASA/IPAC Extragalactic Database (NED), which is operated by the Jet Propulsion Laboratory, California Institute of Technology, under contract with the National Aeronautics and Space Administration.
This paper has been partially supported with grants from Consejo Nacional de 
Investigaciones Cient\'ificas y T\'ecnicas (PIP 11220130100365CO) Argentina, Fondo para la Investigación Científica y Tecnológica (FonCyT, PICT-2017-3301), and Secretar\'ia de Ciencia y Tecnolog\'ia, Universidad Nacional de C\'ordoba, Argentina.
\end{acknowledgements}


\end{document}